\documentclass[aps,prx,reprint,superscriptaddress]{revtex4-1}

\usepackage{graphicx}
\usepackage{dcolumn}
\usepackage{amsmath, amscd, amssymb, amsthm, bm}

\usepackage[dvipsnames]{xcolor}

\usepackage{microtype}  
              
\makeatletter
\def\textless{\afterassignment\textless@\let\next= }
\def\textless@#1#{\@nameuse{textless@#1}}
\def\textless@sub#1#2/sub#3{%
  \ensuremath{_{\let\textless\relax#2}}%
  \egroup 
}
\def\textless@sup#1#2/sup#3{%
  \ensuremath{^{\let\textless\relax#2}}%
  \egroup 
}
\makeatother

\begin{document}

\title{Magneto-Elastic Coupling to Coherent Acoustic Phonon \\ 
Modes in Ferrimagnetic Insulator GdTiO$_3$}

\author{D. Lovinger}
\affiliation{Department of Physics, University of California, San Diego, La Jolla, CA, 92093}
\author{E. Zoghlin}
\affiliation{Materials Department, University of California, Santa Barbara, CA, 93106}
\author{P. Kissin}
\affiliation{Department of Physics, University of California, San Diego, La Jolla, CA, 92093}
\author{G. Ahn}
\affiliation{Department of Physics, Hanyang University, Seoul, South Korea, 04763}
\author{K. Ahadi}
\affiliation{Materials Department, University of California, Santa Barbara, CA, 93106}
\author{P. Kim}
\affiliation{Department of Physics, University of California, San Diego, La Jolla, CA, 92093}
\author{M. Poore}
\affiliation{Department of Physics, University of California, San Diego, La Jolla, CA, 92093}
\author{S. Stemmer}
\affiliation{Materials Department, University of California, Santa Barbara, CA, 93106}
\author{S. J. Moon}
\affiliation{Department of Physics, Hanyang University, Seoul, South Korea, 04763}
\author{S. D. Wilson}
\affiliation{Materials Department, University of California, Santa Barbara, CA, 93106}
\author{R. D. Averitt}
\affiliation{Department of Physics, University of California, San Diego, La Jolla, CA, 92093}

\date{\today}

\begin{abstract}
In this work we investigate single crystal GdTiO$_{3}$, a promising candidate material for Floquet engineering and magnetic control, using ultrafast optical pump-probe reflectivity and magneto-optical Kerr spectroscopy. GdTiO${}_{3}$ is a Mott-Hubbard insulator with a ferrimagnetic and orbitally ordered ground state (\textit{T${}_{C}$} = 32 K). We observe multiple signatures of the magnetic phase transition in the photoinduced reflectivity signal, in response to above band-gap 660 nm excitation. Magnetic dynamics measured via Kerr spectroscopy reveal optical perturbation of the ferrimagnetic order on spin-lattice coupling timescales, highlighting the competition between the Gd${}^{3+}$ and Ti${}^{3+}$ magnetic sub-lattices. Furthermore, a strong coherent oscillation is present in the reflection and Kerr dynamics, attributable to an acoustic strain wave launched by the pump pulse. The amplitude of this acoustic mode is highly dependent on the magnetic order of the system, growing sharply in magnitude at \textit{T${}_{C}$}, indicative of strong magneto-elastic coupling. The driving mechanism, involving strain-induced modification of the magnetic exchange interaction, implies an indirect method of coupling light to the magnetic degrees of freedom and emphasizes the potential of GdTiO${}_{3}$ as a tunable quantum material.

\end{abstract}

\maketitle


\section{\label{sec:introduction}Introduction}

The rare-earth titanates (unit formula RTiO${}_{3}$, where R is a rare-earth ion) are a class of complex materials with strongly correlated spin, orbital, and lattice degrees of freedom. They are 3\textit{d${}^{1}$} compounds with a single \textit{d-}orbital electron occupying the Ti${}^{3+}$ \textit{t${}_{2g}$} orbital, whose degeneracy is broken by strong crystal field splitting \cite{Mochizuki2004b}. This presents an opportunity to study a strongly correlated system in relative simplicity, which nonetheless exhibits rich physics and interesting properties. The perovskite titanates, for example, are Mott-Hubbard (MH) insulators with interconnected orbital and spin order \cite{Tokura1992a,Turner1980a,Okimoto1995a,Itoh1999a,Mochizuki2001c}. Of particular interest is the complex magnetic phase diagram for this class of materials, with a magnetic ground state that varies from ferrimagnetic to antiferromagnetic as a function of the rare-earth ion size and subsequent change in Ti-O-Ti bond angle \cite{Mochizuki2004b,Mochizuki2000a}. Various theories have attempted to explain the magnetic order in titanates \cite{Itoh1999a,Zhou2005a,Takubo2010a,Pavarini2004}, all of which highlight the need to consider the roles of structure and electronic correlation to understand the complexity embodied in the magnetic phase diagram. 

Common to all descriptions of magnetism in the perovskite titanates is the defining role of the lattice and its distortion. It has been argued, for example, that the degree of GdFeO${}_{3}$ distortion and changes to the Ti-O-Ti bond angle directly modify the exchange interaction which, in turn, determines the magnetic order \cite{Mochizuki2000a,Mochizuki2001b}. More recent results emphasize the importance of orbital order in determining the magnetic order. In particular, the direct coupling between the orbital order and lattice, rather than the orthorhombic distortion, contributes most strongly to the ground state \cite{Komarek2007a,Zhang2013a,Varignon2017a}. Whether it is particular structural distortions or more generalized Jahn-Teller distortions, and regardless of the role of orbital ordering, it is clear that magnetic order in titanates is highly dependent on the lattice.

\begin{figure*}
  \centering
  \includegraphics[width=0.9\textwidth]{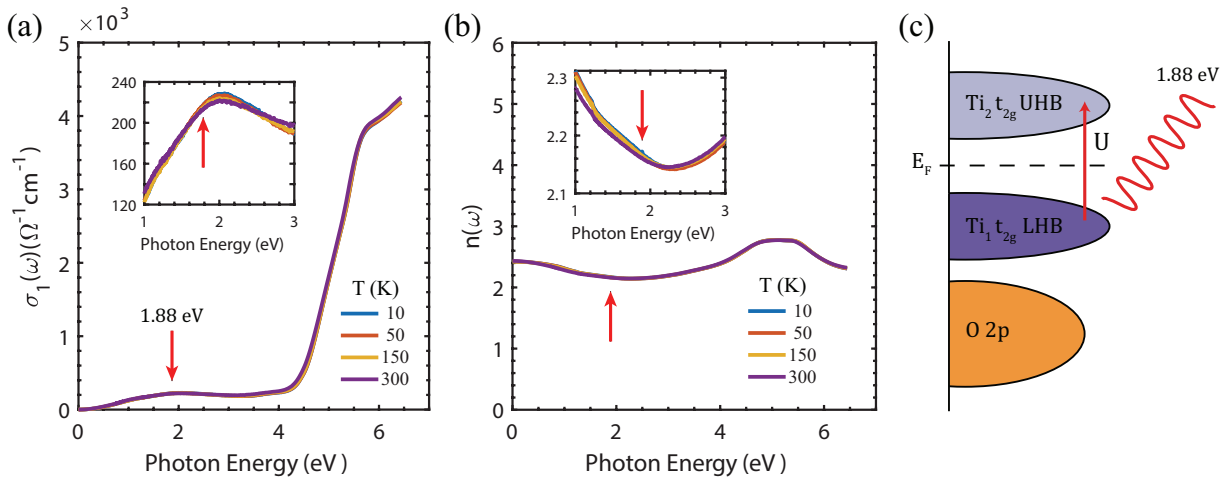}
  \caption[(a) Optical conductivity and (b) index of refraction of GTO/LSAT thin film as a function of photon energy and temperature.]
  {(a) Optical conductivity and (b) index of refraction of GTO/LSAT thin film as a function of photon energy and temperature. Red arrows indicate the 1.88 eV pump/probe energy. The weak feature in $\sigma_1$ at 2 eV corresponds to the MH gap, while the steep feature near 5 eV arises from O$_{2p}$ to Ti$_{3d}$ and Gd$_{4f}$ charge transfer transitions. (c) Depiction of 1.88 eV laser excitation, corresponding to intersite Ti \textit{3d-3d} transition across the MH gap. \index{gto1_optical_constants}}
  \label{fig:gto1_optical_constants}
\end{figure*}

In this work we study GdTiO${}_{3}$ (GTO), a titanate with an orthorhombic perovskite-type unit cell and relatively large GdFeO${}_{3}$-type distortion. GTO lies just within the ferromagnetic (FM) region of the phase diagram. The proximity to the FM-AFM transition makes GTO particularly sensitive to the effect of structural changes on the magnetism \cite{Komarek2007a}. Below the critical temperature \textit{T${}_{C}$} = 32 K it is ferrimagnetically (fM) ordered: the Ti${}^{3+}$ spins are aligned ferromagnetically along the \textit{c}-axis and coupled antiferromagnetically to the Gd sublattice \cite{Turner1980a,Zhou2005a,Komarek2007a}. The magnetism saturates at 6 $\mu_{B}$ (7 $\mu_{B}$ Gd -- 1 $\mu_{B}$ Ti) in a relatively small field of $\mathrm{\sim}$0.1 T, with no discernable hysteresis \cite{Amow2000a}. The magnetocrystalline anisotropy is small, with the \textit{a}-axis as the hard magnetization axis and the \textit{b-c} plane nearly isotropic. The fM order is accompanied and mediated by (\textit{yz, zx, yz, zx})-type orbital order, a result of inter-atomic hybridization between the t\textit{${}_{2g}$} and e\textit{${}_{g}$} orbitals \cite{Mochizuki2004b}. 

The present work on GTO is motivated not only by the relative simplicity of the system and rich interconnected order, but also the potential for Floquet engineering and ultrafast control of magnetism. Liu \textit{et al.} explored the Mott insulating titanates as a candidate for tuning the spin-orbital Floquet Hamiltonian and subsequent modification of the spin exchange interaction using light \cite{Liu2018a}. Meanwhile, Khalsa \textit{et al.} suggest direct excitation of a GTO mid-IR active phonon mode to transiently modify the exchange interaction and switch the ground state from FM to AFM on ultrafast timescales \cite{Khalsa2018b}. A similar experiment utilizing phononic control has been proposed by Gu \textit{et al.} in other titanates \cite{Gu2018}. 

While the conditions of our experiment lie outside the regimes discussed above, we do observe strong coupling between light, the lattice, and the sample magnetism. Time-resolved pump-probe and magneto-optical Kerr effect (MOKE) measurements tuned to $\mathrm{\sim}$1.88 eV, just above the bandgap, allow us to measure the evolution of photoexcited states on femtosecond -- picosecond timescales. We observe multiple signatures of the magnetic phase transition in the photoinduced reflectivity signal, as well as optical perturbation of the fM order on spin-lattice coupling timescales in the MOKE signal. In addition, an acoustic phonon mode is present in both signals, whose amplitude is highly coupled to the magnetic order. This implies strong magneto-elastic coupling through transient, strain-induced modification of the exchange interaction, connecting the lattice and magnetic degrees of freedom and indicating that the exchange interaction is tunable on ultrafast timescales. 

\section{\label{sec:methods}Methods}

Single crystal and thin film samples of GdTiO${}_{3}$ were investigated. The photoinduced reflectivity signal in both is extremely similar and the following work, except for the measurement of the optical constants, was performed on a single crystal sample. For comparison, the thin film time-resolved reflectivity data is presented in SM I \cite{SMbib}. GdTiO${}_{3}$ thin films ($\mathrm{\sim}$20 nm) were grown on a (001)(La${}_{0.3}$Sr${}_{0.7}$)(Al${}_{0.65}$Ta${}_{0.35}$)O${}_{3}$ (LSAT) substrate by hybrid molecular beam epitaxy \cite{Moetakef2013a}. GdTiO${}_{3}$ bulk single crystals were grown by high pressure laser floating zone method \cite{Schmehr2019}. A small fraction of the crystal rod was cut and polished to optical quality, with \textit{bc-}axis in plane and \textit{a-}axis out of plane. Powder X-ray diffraction measurements indicate extremely high quality crystals with no notable impurity peaks and lattice parameters at 5.393, 5.691, 7.664 {\AA} for \textit{a},\textit{b}, \textit{c}-axis \cite{Schmehr2019}, well matched to literature values \cite{Komarek2007a}. Magnetization measurements indicate no visible hysteresis and a saturation moment of 6 $\mu_B/$FU.

To determine the optical conductivity and index of refraction, frequency-dependent reflectivity spectra \textit{R}($\omega$) in the photon energy region between 3 meV and 85 meV were measured by using a Bruker VERTEX 70v Fourier transform spectrometer. The GdTiO${}_{3}$ thin film was mounted in a continuous liquid helium flow cryostat. We used two spectroscopic ellipsometers (IR-VASE Mark II and M-2000, J. A. Woollam Co.) for obtaining the complex dielectric constants $\epsilon(\omega) = \epsilon_1(\omega) + i\epsilon_2(\omega)$ in the energy range from 60 meV to 0.75 eV and 0.75 eV to 6.4 eV, respectively. The optical conductivity of the GdTiO${}_{3}$ film was obtained by two-layer model fit employing Drude-Lorentz oscillators for optical response of each layer \cite{Kuzmenko2005}. 

Ultrafast optical pump-probe reflectivity measurements ($\Delta$R/R) are performed using a 1040 nm 200 kHz Spectra-Physics Spirit Yb-based hybrid-fiber laser coupled to a non-colinear optical parametric amplifier. The amplifier produces $\mathrm{\sim}$20 fs pulses centered at 660 nm (1.88 eV), which are split, cross-polarized (pump \textit{s}-polarized, probe \textit{p}), and used as degenerate pump and probe beams. The pump is aligned along the \textit{b}-axis of the GTO crystal. This excitation corresponds to an intersite Ti \textit{3d} -- \textit{3d} transition across the Mott-Hubbard gap, shown in Fig. \ref{fig:gto1_optical_constants}(c). A moderate pump fluence of $\mathrm{\sim}$100 $\mu$J/cm${}^{2\ }$is used to minimize sample heating ($\mathrm{\sim}$4 K at 10 K), ensuring we are in the linear excitation regime. 

Time-resolved magneto-optical Kerr spectroscopy is used to probe the magnetization dynamics. The same optical system described above is used here, including laser energy, fluence, and magneto-optical cryostat (Quantum Design OptiCool). The photoinduced Kerr rotation ($\mathrm{\Delta }{\mathrm{\theta }}_K$) is measured using balanced photodiodes in the polar Kerr geometry at near-normal incidence, in a continuously variable external magnetic field (0 -- 7 T), with the pump polarized along the \textit{b}-axis and Kerr probe polarized along the \textit{c}-axis of the crystal. The magnetic field is applied normal to the sample surface, along the \textit{a-}axis of the crystal, resulting in a Kerr signal proportional to the out-of-plane z-component of the photoinduced change in magnetization $\mathrm{\Delta }M_z$. In order to eliminate non-magnetic contributions to the signal and ensure we are measuring genuine spin dynamics, we take the difference of the Kerr signal at various positive and negative applied fields: $\mathrm{\Delta }{\mathrm{\theta }}_K=\mathrm{\Delta}\theta \left(+M\right)-\mathrm{\Delta}\theta \left(-M\right)$ (see SM IV for details) \cite{SMbib, Jones1941a, Veis2009, KLIGER199059, Nyvlt1996a, Wang2012Ch}.

\section{\label{sec:results}Experimental Results}

The temperature-dependent optical conductivity of thin-film GdTiO$_3$ is shown in Fig. \ref{fig:gto1_optical_constants}(a) for photon energies ranging from 3 meV to 6.5 eV, and for temperatures from 10 K to 300 K. A weak feature is present, centered at 2 eV, corresponding to the Mott-Hubbard gap. Apart from weak thermal broadening with increasing temperature, the peak at 2 eV is nearly temperature independent. While early studies of GTO measured a MH gap of 0.2 -- 0.7 eV \cite{Crandles1992}, more recent photoluminescence and DFT/DFT+U results place the gap closer to 1.8 -- 2 eV \cite{Bjaalie2015a}. The small peak in the optical conductivity spectrum at 2 eV measured here supports these recent findings. At much higher energies we observe a significant increase in the optical conductivity. The features near 5 eV correspond to O${}_{2p}$ to Ti${}_{3d}$ and Gd${}_{4f}$ transitions. Fig. \ref{fig:gto1_optical_constants}(b) shows the index of refraction in the same energy range, which remains relatively constant as a function of temperature.

\begin{figure}[t] 
  \centering
  \includegraphics[width=\linewidth]{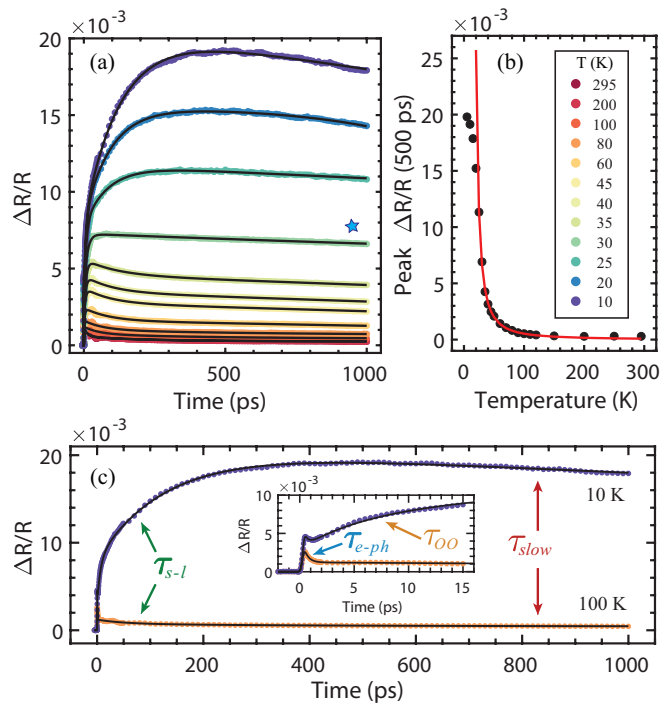}
  \caption[(a) Photoinduced differential reflectivity signal $\Delta R/R$ at all measured temperatures, from 10 K to 295 K.]
  {(a) Photoinduced differential reflectivity signal $\Delta R/R$ at all measured temperatures, from 10 K to 295 K, taken on a single crystal GdTiO$_3$ sample. The black curves are exponential fits to the data, of the form given in Eq. 1. The blue star indicates the data curve taken at $T_C$. (b) $\Delta R/R$ values at 500 ps, an approximation of the peak signal at all temperatures. The red line is a power law fit, commonly seen in systems undergoing a second-order magnetic phase transition. (c) Representative pump-probe scans at 10 K and 100 K, indicating the various timescales involved in the recovery process (see Eq. \ref{eq:dRR}). \index{gto2_drr}}
  \label{fig:gto2_drr}
\end{figure}

The time-dependent photoinduced change in reflectivity $\mathrm{\Delta }R/R$ for a GdTiO$_3$ single crystal is shown in Fig. \ref{fig:gto2_drr}(a), for all measured temperatures between 10 -- 295 K (legend on Fig. \ref{fig:gto2_drr}(b)). The black lines represent exponential fits to the data as described below. The photoinduced change in $\mathrm{\Delta }R/R$ is positive; following laser excitation a non-equilibrium electron population is established in $\mathrm{\sim}$500 fs, which then exchanges energy and equilibrates with the spin and lattice subsystems through various pathways, each with a characteristic timescale. This is visible as the slower, multi-component exponential relaxation. As the temperature is decreased from 295 K the signal amplitude increases, recovery dynamics slow, and two additional features emerge. The first is a delayed rise time, corresponding to a further departure from equilibrium in the first $\mathrm{\sim}$15 ps, emerging below \textit{T} = 100 K. Second, there is a crossover point visible at delay times of $\mathrm{\sim}$200 ps where recovery dynamics flatten and reverse direction to become an additional rise time. This occurs precisely as the ferrimagnetic ordering temperature \textit{T${}_{C}$} = 32 K is crossed (marked by a blue star), indicating that magnetization dynamics manifest in the differential reflectivity signal. 

To further investigate the temperature dependence of the reflectivity signal, we plot the peak signal amplitude (at 500 ps) in Fig 2(b). The behavior here is distinctive, not uncommon in materials undergoing a second-order magnetic phase transition. The red curve represents a power-law fit to the data, of the form $A=A_0t^{-w}$, where \textit{w} is the critical exponent and \textit{t} is the reduced temperature $t=\ \frac{\left|T-T_C\right|}{T_C}$. The value of \textit{w} depends upon the symmetry and universality class of the magnetic transition. Our fit produces a critical exponent $w = 1.28\pm 0.02$. This very nearly matches the critical behavior predicted by dynamical scaling theory for the 3D Ising model, which yields $w \approx 1.32$ \cite{Hohenberg1977a,Wang1995,Pelissetto2002} (further detailed in SM II) \cite{SMbib, Stanley1987b, Hohenberg1977a, Wang1995, Pelissetto2002, Campostrini2002, Hinton2015a}. While this is an indirect method of measuring critical dynamics and is not intended to be a rigorous analysis, it is clear that the peak reflectivity follows power-law behavior as expected at a magnetic phase transition. The results suggests that there is indeed a magnetic contribution in the $\Delta R/R$ signal. Additionally, the qualitative form of the peak amplitude vs temperature follows that of the temperature dependent magnetic susceptibility in bulk GTO \cite{Amow2000a}, and the magnetization M in films \cite{Moetakef2012a, Schmehr2019}. While by no means conclusive, the universal scaling behavior and agreement with thermal magnetization does strongly suggest that the $\mathrm{\Delta }R/R$ signal measured, particularly at longer times (500+ ps), is sensitive to spin dynamics.

\begin{figure}[t] 
  \centering
  \includegraphics[width=\linewidth]{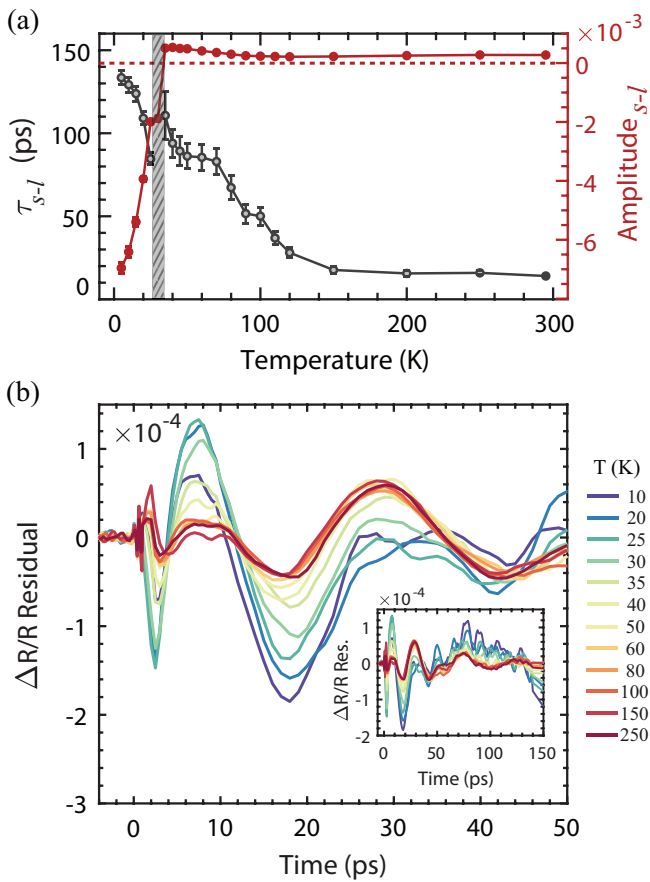}
  \caption[(a) Time constant (black) and amplitude (red) of the spin-lattice coupling term, extracted from exponential fits to the $\Delta R/R$ data.]
  {(a) Time constant (black) and amplitude (red) of the spin-lattice coupling term, extracted from exponential fits to the $\Delta R/R$ data. The vertical gray section indicates the fM transition region $T_C = 32$ K, where the lifetime is too long to measure (see 30 K curve marked by a blue star in Fig. \ref{fig:gto2_drr}(a)). The red dashed line depicts zero amplitude and clarifies the crossover region. (b) Coherent acoustic phonon response, isolated by subtracting the exponential fits from the $\Delta R/R$ data. The inset shows the $\Delta R/R$ residual to a longer delay time of 150 ps, where higher frequency components emerge. \index{gto3_drr_res}}
  \label{fig:gto3_drr_res}
\end{figure}

To substantiate these claims, we quantitatively analyze the full time-dependent response. Below 100 K, the dynamics can be fit by a sum of four exponentials with a constant offset, of the form:

\begin{multline}
\label{eq:dRR} 
\mathrm{\Delta }R/R\left(t\right) = \color{MidnightBlue}A_{e-ph}e^{-t/{\tau }_{e-ph}} \color{black}+ \color{Bittersweet}{A_{OO}e^{-t/{\tau }_{OO}}} \color{black}+ \\
\color{OliveGreen}A_{s-l}e^{-t/{\tau }_{s-l}} \color{black}+ \color{red}A_{slow}e^{-t/{\tau }_{slow}} \color{black}+ C, 
\end{multline} 

\noindent shown as black lines in Fig. \ref{fig:gto2_drr}(a). Not listed is an additional error function term, which describes the initial step-like rise dynamics at $t = 0$. A visual representation of the various timescales is shown in Fig. \ref{fig:gto2_drr}(c) for two temperatures. After excitation the dynamics follow a general trend; there is a very fast initial recovery, ${\tau }_{e-ph}$ on the order of $\mathrm{\sim}$500 fs, followed by an intermediate term ${\tau }_{OO}$ on the order of  2 -- 8 ps, both of which are clearly visible in the inset of Fig. \ref{fig:gto2_drr}(c). Note that ${\tau }_{OO}$ is an additional rise time which vanishes at higher temperatures, the full dynamics fitting to only 3 exponentials (i.e. above \textit{T${}_{C}$}). This is followed by a slower term ${\tau }_{s-l}$ on the order of 100's of picoseconds, and a final much slower recovery ${\tau }_{slow}$. A careful inspection of the reflectivity data also indicates the presence of small oscillations about the black fitted curves, which we discuss below.

These measured timescales are well separated and can be attributed to distinct physical processes. The initial pump pulse excites an intersite Ti \textit{3d-3d} transition. This directly creates a population of hot carriers which thermalize via electron-electron (e-e) scattering, then subsequently exchange energy with the lattice, orbital, and spin degrees of freedom. We focus on the spin-lattice coupling process here, with a full discussion of the remaining processes and time constants in SM III \cite{SMbib, Qi2012,Wall2009,MiyasakaNakamura2006, Mochizuki2004b, Mochizuki2001b, Furukawa1996a,Furukawa1997a,Itoh2001a, Tomimoto2003, Gossling2008a, Novelli2012a}.

The most relevant component of the $\mathrm{\Delta }R/R$ signal is the third fitted exponential, ${\tau }_{s-l}$, attributed to spin-lattice coupling and shown in Fig. \ref{fig:gto3_drr_res}(a). This term has a characteristic timescale of 10 -- 140 ps, excluding the region at the magnetic phase transition temperature \textit{T${}_{C}$ }= 32 K where the lifetime grows too long to accurately measure. This critical region is visible as a flattening of the $\Delta$R/R recovery at 30 K, indicated by the blue star in Fig. \ref{fig:gto2_drr}(a). Before the onset of this third recovery term ${\tau }_{s-l}$, the $\Delta$R/R signal reveals dynamics indicative of electron-lattice equilibration. It follows that this longer lifetime is related to equilibration of the spin subsystem with the lattice. The time constant measured, on the order of 100 ps in the magnetic phase, is consistent with spin-lattice coupling in other magnetic insulators \cite{Wang2005a,Vaterlaus1991c,Kimel2002a}. The characteristic time is relatively constant in the paramagnetic phase until $150-200$ K, where it begins to slowly increase. This corresponds to the onset temperature ($\mathrm{\sim}$180 K) of spin-spin coupling between the Gd${}^{3+}$ and Ti${}^{3+}$ ions \cite{Zhou2005a}. Closer to 100 K ${\tau }_{s-l}$ further increases, indicating the onset of short-range fM spin correlations. This is also apparent in the increase in amplitude at this temperature. Finally, as \textit{T${}_{C}$ }is crossed (dark gray region) we see evidence of the second-order ferrimagnetic phase transition as the time constant diverges and amplitude switches sign. The now-negative amplitude implies an additional rise time in the signal; as energy is transferred to spins and the ferrimagnetic order is disrupted, the system is brought further out of equilibrium. In the paramagnetic phase there is no long-range spin order to disrupt, such that spin-lattice thermalization manifests as a simple recovery to equilibrium. The critical behavior, amplitude reversal, timescale, and temperature dependence of the ${\tau }_{s-l}$ component all suggest that we are measuring spin-lattice coupling on a timescale of $\mathrm{\sim}$100 ps, and that it is highly sensitive to the onset of magnetic order.

The final interesting feature of the $\mathrm{\Delta }R/R$ data is a slow coherent oscillation, prominent at early times. By subtracting the exponential fits at each temperature we can extract the oscillatory component, plotted in Fig. \ref{fig:gto3_drr_res}(b). The result is peculiar -- we observe a low-frequency phonon mode which grows in amplitude and becomes chirped, slowing down and redshifting as it propagates. The oscillation period (on the order of 20 ps), suggests an acoustic strain wave launched by the pump pulse which propagates through the crystal \cite{C.ThomsenH.T.GrahnH.J.Maris1986a}. The probe beam reflected from the sample surface interferes with a portion reflected from the strain wave boundary, resulting in an oscillatory signal. The temperature dependence of this mode is striking -- the amplitude is relatively constant at high temperatures, then grows sharply precisely at the fM phase transition temperature. Though it appears to be an acoustic mode, it is also clearly coupled to the magnetic order. This suggests strong magneto-acoustic coupling, tying the dynamics of the magnetic subsystem to the transiently strained lattice. 

\begin{figure}[t] 
  \centering
  \includegraphics[width=\linewidth]{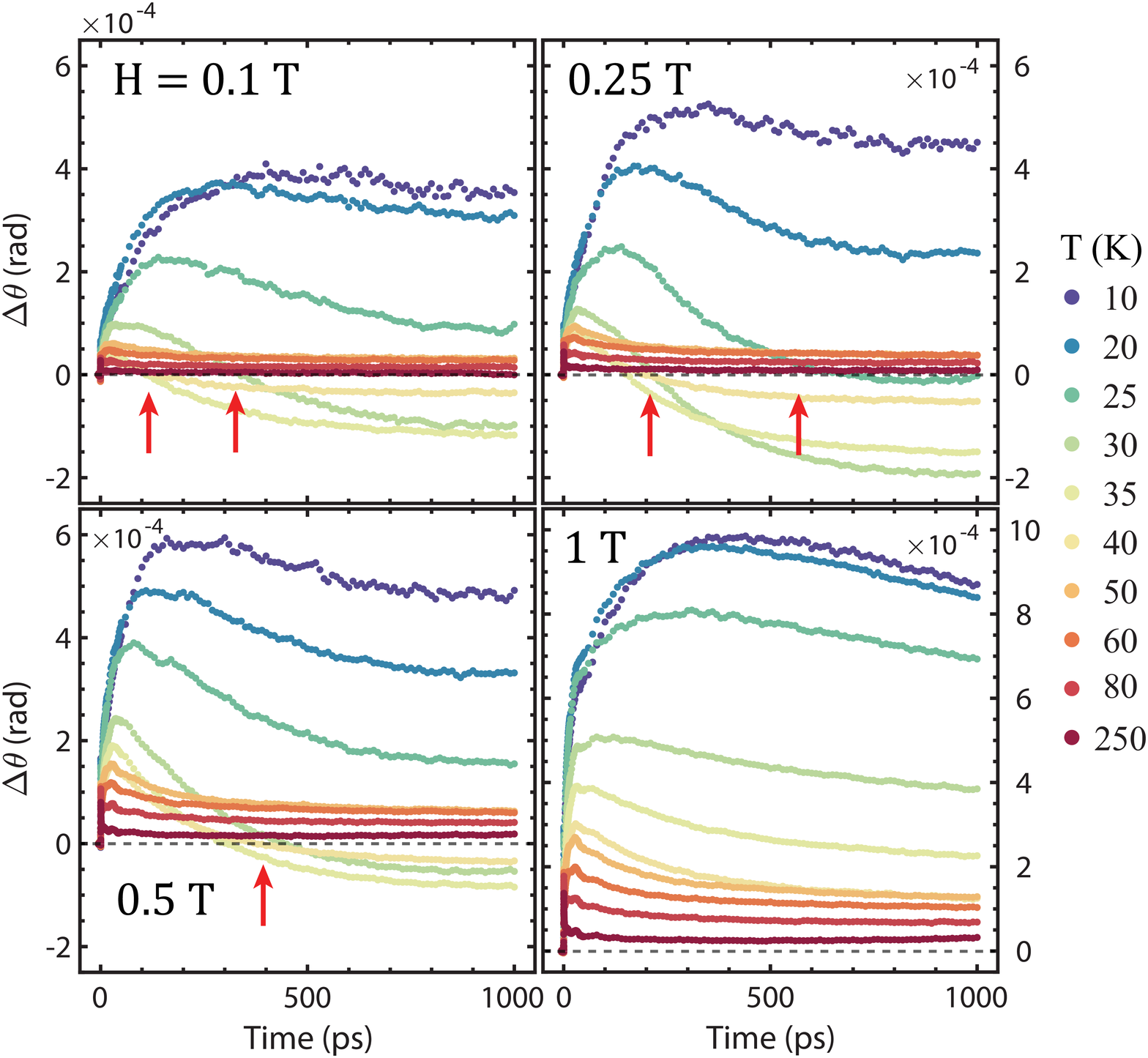}
  \caption[Time-resolved Kerr dynamics at various magnetic fields, recorded as the difference between the MOKE signal in opposing field directions.]
  {Time-resolved Kerr dynamics at various magnetic fields, recorded as the difference between the MOKE signal in opposing field directions $\Delta \theta = \Delta \theta(+H) - \Delta \theta(-H)$. This is a measure of the photoinduced change in the out-of-plane magnetization $\Delta M_z$. Red arrows indicate the crossover to negative values of $\Delta \theta$. \index{gto4_kerr}}
  \label{fig:gto4_kerr}
\end{figure}

To gain further insight into the magnetization dynamics and the influence upon acoustic phonon propagation, we utilize time-resolved magneto-optical Kerr effect (MOKE) spectroscopy. Fig. \ref{fig:gto4_kerr} presents the photoinduced Kerr rotation $\Delta\theta$, proportional to the change in out-of-plane (\textit{a}-axis) magnetization $\mathrm{\Delta}M_z$, for all temperatures and four fields between 0.1 -- 1 T. For details of the analysis, see SM IV \cite{SMbib, Jones1941a, Veis2009, KLIGER199059, Nyvlt1996a, Wang2012Ch}. Additional static Kerr rotation measurements are presented in SM V \cite{SMbib}. At lower field strengths, the Kerr signal reveals a quick rise in the photoinduced out-of-plane magnetization $\mathrm{\Delta }M_z$, followed by a reduction and change in sign of $\mathrm{\Delta }M_z$. This can be interpreted as a pump-induced increase and subsequent decrease in the net out-of-plane magnetic moment, but not necessarily a reversal of the total magnetic moment. There are two primary components to the Kerr signal, one positive (growing in $\mathrm{\sim}$100 ps), and one negative (growing in slower, $\mathrm{\sim}$100 -- 500 ps). These dynamics are slow and long-lived, as expected in magnetic insulators like GTO due to the localized nature of quasiparticles \cite{Kimel2002a}. To describe the temperature dependence of the signal, we focus on lower field strengths $H=0.1-0.5$ T. At high temperature, in the paramagnetic phase, the Kerr signal is weak and indicates the lack of long-range magnetic order. As the temperature is lowered there is an increase in the photoinduced rotation, with a clear negative signal emerging below \textit{T${}_{C}$}. This negative component is largest and appears at earlier delays right at the transition temperature (\textit{T${}_{C}$} = 32 K). With decreasing temperature, the crossover to negative values of $\mathit{\Delta}\theta $ occurs at later times. Well below \textit{T${}_{C}$}, in the strongly ordered phase, the signal remains positive at all time delays.

We also observe a significant field dependence in the data. The maximum signal amplitude at all temperatures increases with increasing field. In addition, the negative amplitude component is most pronounced at 0.25 T, decreasing in amplitude at higher fields and vanishing entirely by 1 T. At this high field, we note that the photoinduced magnetization dynamics look qualitatively similar to the photoinduced reflectivity signal $\mathrm{\Delta }R/R$ shown in Fig. \ref{fig:gto2_drr}(a). In the $\mathrm{\Delta }R/R\ $data, the measured signal is primarily the result of Ti sublattice dynamics due to the 1.9 eV intersite Ti-Ti excitation, and is dominated by Ti spin dynamics: the spin-lattice and spin relaxation terms (${\tau }_{s-l}$ and ${\tau }_{slow}$). It follows that the MOKE signal measured at 1 T is primarily a measure of Ti spin dynamics due to its similarity with the $\Delta R/R$ signal. Fits to the 1 T MOKE data support this, yielding a component with a timescale of 100 -- 200 ps and a very similar temperature dependence when compared to ${\tau }_{s-l}$ extracted from the $\mathrm{\Delta }R/R\ $data (see SM VI for details) \cite{SMbib}. As the field is lowered from 1 T, the magnetization dynamics must be increasingly influenced by the Gd spins. The ferrimagnetic nature of GTO, with two competing magnetic sublattices, is key to understanding the observed behavior as we now discuss.

\section{Discussion}
\label{sec:discussion}

\begin{figure}[t] 
  \centering
  \includegraphics[width=0.9\linewidth]{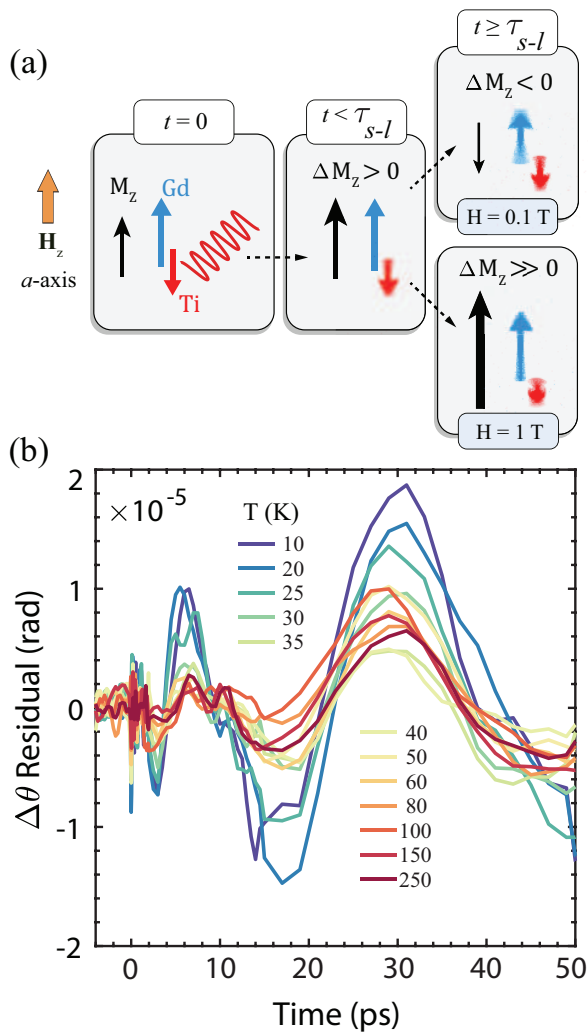}
  \caption[Time-resolved Kerr dynamics at various magnetic fields, recorded as the difference between the MOKE signal in opposing field directions.]
  {(a) Schematic depiction of the spin dynamics. Photoexcitation directly perturbs the Ti spins, which fluctuate and decrease their projection along the applied field H (parallel to \textit{a}-axis, $+z$) in $t < \tau_{s-l}$ ps, increasing the MOKE signal. At longer times: if H and/or magnetic order is weak, induced spin fluctuations and the AFM exchange coupling lowers the projection of Gd spins along \textit{z}. Conversely, if H is larger than the exchange field, at 1 T, Gd does not reorient and no negative component of the signal is observed. (b) Coherent acoustic phonon response, isolated by subtracting the exponential fits from the time-resolved Kerr data (at 1 T applied field). The dynamics appear similar to the $\Delta R/R$ residual, implying a common origin which we attribute to a coherent strain wave launched by the pump pulse. The appearance of this signal in the Kerr response indicates coherent acoustic phonon manipulation of the magnetic order, presumably from exchange modulation. \index{gto5_kerr_residual}}
  \label{fig:gto5_kerr_residual}
\end{figure}

GTO is ferrimagnetic, the Ti and Gd sublattices coupled via an AFM exchange interaction. Gd spins have a significantly larger magnetic moment than Ti, $7\ {\mu }_B$ vs $1\ {\mu }_B$ respectively \cite{Amow2000a}. Below \textit{T${}_{C}$}, at zero field, the two sublattices are aligned into fM domains such that there is no macroscopic moment. As the applied field H along the \textit{a}-axis is increased, spins are rotated to form long-range collinear fM order, with the Gd sublattice aligned parallel to H and Ti anti-parallel. In a field of only 0.1 T saturation is approached, with spins slightly canted from the \textit{a}-axis/H and a net magnetization of $M\approx 5\ {\mu }_B$. With increasing field, canting and spin fluctuations are reduced, increasing the net moment along H. As we approach 1 T, spin fluctuations are minimized and the magnetization becomes saturated at $M\approx 6.0\ {\mu }_B$ \cite{Amow2000a}. The interaction of these two competing magnetic sublattices after photoexcitation will depend on the temperature and applied field and is illustrated in Fig. \ref{fig:gto5_kerr_residual}(a). The 1.9 eV pump pulse directly excites the Ti sublattice, increasing Ti spin fluctuations on timescales $t<{\tau }_{s-l}$. This causes partial reorientation and a decrease in the projection of Ti spins along the Gd moment, corresponding to a rapid increase of $\mathrm{\Delta }M_z$ and a rise in the MOKE signal (i.e. the Ti sublattice magnetization oriented along \textit{-z} is decreased, leading to an overall increase in the net magnetization in the +\textit{z} direction due to the ferrimagnetic order). Various pathways exist which may perturb the Ti spins on such timescales, including spin-orbit coupling \cite{Beaurepaire1998b,Ogasawara2005a} (orbital order is disrupted in $t<8$ ps, see SM III) \cite{SMbib, Mochizuki2004b, Mochizuki2001b, Furukawa1996a,Furukawa1997a,Itoh2001a, Tomimoto2003, Gossling2008a, Novelli2012a}, and exchange modification, discussed below. Subsequently, energy is transferred to the Gd sublattice through spin-lattice thermalization on a timescale $t\ \ge {\tau }_{s-l}$. The spin-lattice coupling timescale of 100 -- 200 ps measured from fits to the $\mathrm{\Delta }R/R$ and MOKE data corresponds to the timescale on which the MOKE signal changes sign, indicating the delayed contribution of Gd spins to the signal. Such ultrafast magnetic sublattice dynamics, albeit with different mechanisms, have been discussed in a variety of materials, including those with similar rare-earth/transition metal correlations \cite{Radu2011, Chen2019}.

The behavior that follows is field-dependent. At low fields, at times on the order of $\tau_{s-l}$, the additional heat transfer and the strong AFM exchange coupling between Gd spins and partially-reoriented Ti spins causes a reduction in the Gd moment along the field direction. This is seen as the negative component, decreasing the signal on spin-lattice timescales until the net $\mathrm{\Delta }M_z$ is negative. At higher field strengths, the applied field locks Gd moments in place parallel to the field direction, minimizing fluctuations. After photoexcitation, the net magnetization only increases as the anti-parallel Ti spins fluctuate and partially reorient. This is true also at low temperatures where the magnetic order is more firmly established, and explains why $\mathrm{\Delta }M_z$ goes negative only in the weakly ordered state near \textit{T${}_{C}$}. 

\begin{figure*}
  \centering
  \includegraphics[width=0.85\textwidth]{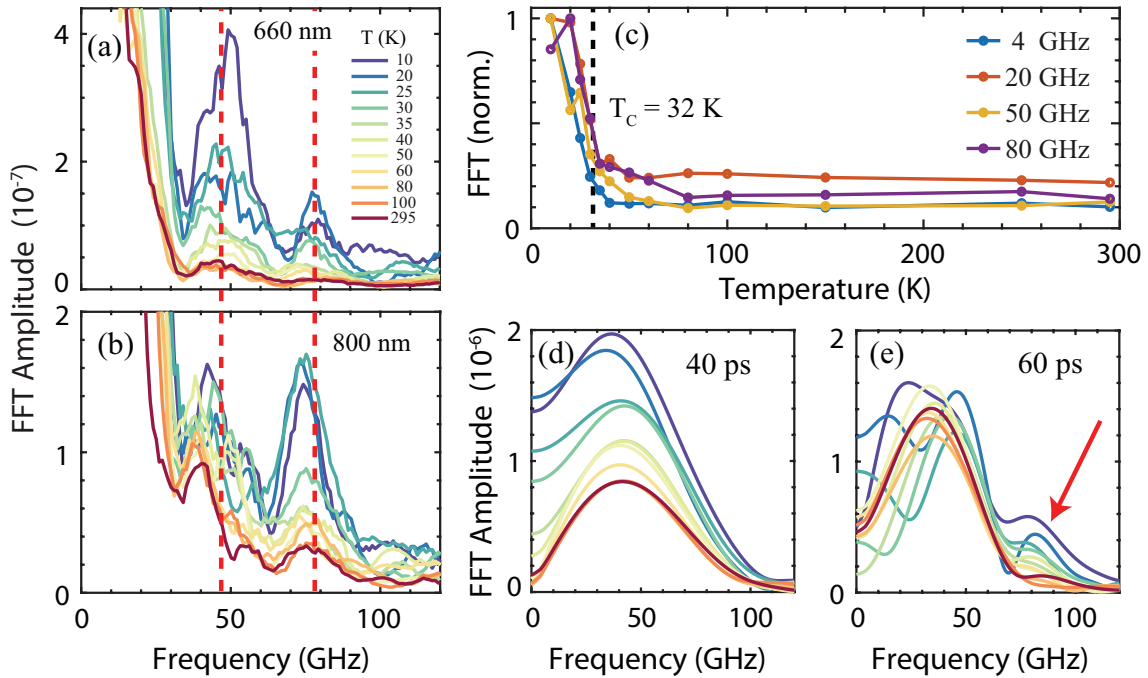}
  \caption[(a) Optical conductivity and (b) index of refraction of GTO/LSAT thin film as a function of photon energy and temperature.]
  {FFT amplitude of the $\Delta R/R$  residual, taken at pump/probe wavelength of 660 nm (a) and 800 nm (b). The red dashed lines indicate the approximate peak positions of the 660 nm FFT. Note the redshift to lower frequency at higher wavelength. (c) The integrated FFT amplitudes at various frequencies as a function of temperature, normalized. (d) The FFT limited to the first 40 ps and (e) 60 ps of the $\Delta R/R$ data. The higher frequency mode emerges only after 40 ps.  \index{gto6_optical_constants}}
  \label{fig:gto6_optical_constants}
\end{figure*}

Finally, we cannot discount the possibility of direct photo-induced modification of the exchange interactions. While the simplest explanation of the MOKE signal involves only heating and spin-lattice coupling, the overall heating is small (no more than $\mathrm{\sim}$4 K at the lowest temperatures at the fluence used). It is therefore not unreasonable to consider more direct electronic changes to the system. The exchange interaction in the titanates is highly dependent on the Ti-O-Ti bond angle and degree of GdFeO${}_{3}$ distortion, as well as the orbital order and occupation \cite{Mochizuki2000a,Mochizuki2001b}. GTO in particular lies on the cusp of the AFM-FM phase boundary, making it especially susceptible to changes in these parameters. Photoexcitation directly disrupts the orbital occupation, which could affect the octahedral distortion and thus the spin exchange interaction. This in turn would provide the drive for reorientation of Ti spins and change in M${}_{z}$ at timescales $t<{\tau }_{s-l}$, and for subsequent perturbation of Gd spins through exchange coupling with Ti. Further calculations of the energy scales of the Ti-Gd exchange field and corresponding timescales are required to confirm this.

 To compare the magnetic dynamics to the $\mathrm{\Delta }R/R$ response, we fit the MOKE data to a series of exponentials similar to Eq. 1 and subtract the fits. Once again, a slow coherent oscillation is revealed, shown in Fig. \ref{fig:gto5_kerr_residual}(b) for the data taken at 1 T. The similarity of the oscillatory Kerr signal to the oscillation in $\mathrm{\Delta }R/R$ is striking -- both phonon modes have the same frequency, same time-dependent redshift, and same temperature dependence, with the amplitude growing rapidly at \textit{T${}_{C}$}. We rule out the possibility of a magnon -- at lower fields there is no change in the frequency of oscillation as we would expect from coherent spin precession (SM VII) \cite{SMbib}. The amplitude is highly field-dependent however, becoming much smaller at lower fields. These observations suggest that the oscillatory mode in the MOKE signal, necessarily a magnetic phenomenon due to the nature of the measurement technique, has the same origin as the oscillatory mode in the $\mathrm{\Delta }R/R$ signal. This is consistent with our interpretation of an acoustic strain wave with strong magneto-elastic coupling. This mechanism has been studied in a variety of ferromagnetic systems, and involves elastic stress modifying the magnetic anisotropy, which exerts a torque on the spins and alters the net magnetization \cite{Weiler2011d,Rossi2005d,Streib2018d}.

To quantify the acoustic phonon response, we show in Fig. \ref{fig:gto6_optical_constants}(a) the FFT of the full $\Delta R/R$ residual, taken from Fig. \ref{fig:gto3_drr_res}(b) (inset). The oscillatory mode with $\mathrm{\sim}$20 ps period featured in Fig. \ref{fig:gto3_drr_res}(b) appears as a strong peak at $\mathrm{\sim}$50 GHz. In this region of interest, it is apparent that there are additional higher frequency modes in addition to the 50 GHz mode. The temperature dependence is also clear; while the FFT amplitude is nearly constant at high temperatures, it grows rapidly upon approaching $T_C$ = 32 K and a higher frequency peak at $\mathrm{\sim}$80 GHz emerges. This again suggests coupling to the magnetic order. Fig. \ref{fig:gto6_optical_constants}(b) applies the same FFT analysis to data taken at an increased pump/probe wavelength of 800 nm. The features are similar, but exhibit a clear redshift as indicated by the red dashed lines. This behavior is consistent with an acoustic strain wave since it arises (for $\Delta R/R$) from interference of the probe with itself. The phonon frequency is wavelength dependent, its form is given by:
\begin{equation}
    f=2nv/\lambda, 
\end{equation}
where \textit{n} is the index of refraction, \textit{v} is the sound velocity, and $\lambda$ is the probe wavelength \cite{C.ThomsenH.T.GrahnH.J.Maris1986a}. As we observe, a higher probe wavelength results in a lower frequency acoustic phonon. Using the measured index of refraction \textit{n} in Fig. \ref{fig:gto1_optical_constants}(b) we can also estimate the sound velocity. At the lower frequency peak near 50 GHz we obtain a sound velocity of  $7.2\times {10}^{3}\ m/s$ and $7.7\times {10}^{3}\ m/s$ for a 660 and 800 nm probe, respectively. This is a very reasonable range for acoustic propagation in solid materials. These results, and the fact that the oscillation frequency does not depend on magnetic field, confirms our classification of the phonon mode as an acoustic strain wave. 

 To more closely examine the link to magnetism, we plot the integrated FFT amplitudes for all frequency peaks in Fig. \ref{fig:gto6_optical_constants}(c). The normalized curves show a striking trend; the amplitude is nearly constant at high temperatures, but sharply increases at or very near to the magnetic ordering transition. The temperature dependence of the FFT amplitudes follows the magnetic order parameter and is remarkably similar to the divergence one expects at a second-order magnetic phase transition. This indicates the presence of magneto-elastic coupling. The acoustic attenuation of sound waves near magnetic phase transitions is well studied, and literature suggests that the attenuation follows power law behavior, similar to our result \cite{Weiler2011d}. In the vicinity of \textit{T${}_{C}$}, energy density and spin fluctuations play the primary role in attenuation. This behavior has been studied in a wide range of magneto-elastically coupled materials, including Ni \cite{Weiler2011d}, CoF${}_{2}$ \cite{Thomson2014b}, and MnF${}_{2}$ \cite{Neighbors1968b}. 

A final interesting feature to note is shown in Fig. \ref{fig:gto6_optical_constants}(d-e), comparing an FFT of the $\mathit{\Delta}R/R$ data limited to the first 40 ps (d) and to the first 60 ps (e) of the scan. This analysis reveals that the high frequency component at 80 GHz begins to emerge only after 40 ps, which is also visible in the time-domain data (Fig. \ref{fig:gto3_drr_res}(b) inset). This timescale is similar to the spin-lattice coupling timescale measured in both $\mathrm{\Delta }R/R$ and MOKE, which ranges from $\mathrm{\sim}$50 -- 150 ps. We have also discussed the spin dynamics following photoexcitation, where Ti spins are immediately perturbed and Gd follows after exchange pathway alterations and spin-lattice thermalization. Given the similar timescales, we suggest that the emergence of the 80 GHz mode indicates the onset of Gd spin dynamics. Roughly 50 ps after photoexcitation the Gd spin subsystem begins thermalizing and fluctuating. This damps the acoustic oscillation and changes the magnetic background. The now higher energy of the Gd spins alters the spin-phonon and magnetostrictive interaction strengths, resulting in a change to the magnetically-coupled elastic parameters of the lattice and a subsequent shift in phonon frequency. 

 A microscopic description of magneto-elastic coupling involves a transient modification of the exchange interaction. As the acoustic wave propagates it modulates the distance between lattice sites and spins. This in turn produces a periodic modification of the exchange interaction between neighboring spins, coupling the acoustic wave to the magnetic order parameters. The result is an attenuation of the acoustic wave in the high-temperature phase where spin fluctuations are large, lessening as spin correlations increase in the low temperature ordered phase. The same mechanism decreases acoustic attenuation, increasing the phonon amplitude, in an applied magnetic field as observed in our MOKE signal. This has been described by an approximate analytical theory \cite{Neighbors1968b,Bennett1967b,Ghatak1972b}, which generally predicts maximal acoustic damping at the critical point and a MHz frequency shift in the ordered phase. We observe that the damping is consistently large throughout the high temperature paramagnetic phase, and we do not observe such a frequency shift with temperature. In our experiment, however, a MHz frequency shift is too small to be observed, and the dynamics at picosecond timescales are strongly coupled to out-of-equilibrium degrees of freedom that will affect the acoustic wave propagation and attenuation in other unanticipated ways. 

 The phonon behavior we observe undoubtedly suggests a strong coupling of the lattice to the magnetic order in GdTiO${}_{3}$. Furthermore, the mechanism implies transient exchange modification on an ultrafast timescale. These conclusions are not without precedence. Ultrafast magneto-elastic coupling has been demonstrated by Bigot \textit{et al.}, for example, in Ni thin films \cite{Kim2012b}, with experiments going so far as to control the magnetic precession through acoustic pulses \cite{Kim2015b}. Kimel \textit{et al.} have shown optical quenching of magnetic order through phonon-magnon coupling in FeBO${}_{3}$ \cite{Kimel2002a} and Nova \textit{et al.} have shown that Mid-IR and THz excitation resonant with specific lattice modes is able to drive collective spin precession \cite{Nova2017b}. Our work represents another potential method of using light to indirectly alter the magnetic degrees of freedom on ultrafast timescales, through coupling to an acoustic phonon mode. 
 
 \section{Conclusion}
 \label{sec:conclusion}
 
We have used a multi-modal approach, consisting of time-resolved photoinduced reflectivity and magneto-optical Kerr (MOKE) spectroscopy, to study magneto-elastic coupling in the ferrimagnetic insulator GdTiO${}_{3}$. We observe multiple, clear signatures of the ferrimagnetically ordered phase at \textit{T${}_{C}$} = 32 K in both signals, and measure spin-lattice thermalization timescales ${\tau }_{s-l}$ on the order of 100 picoseconds, as might be expected in a magnetic insulator.

From the MOKE signal we observe long-lived spin dynamics and optical perturbation of the ferrimagnetic order. This includes a change in sign of the photoinduced magnetization on the same timescale as spin-lattice coupling. The ferrimagnetic nature of GTO, with two magnetic sublattices coupled antiferromagnetically, is responsible. Photoexcitation at 660 nm directly perturbs the Ti moments, increasing fluctuations and causing a partial reorientation and decrease in the projection of Ti spins along the Gd moment. This is measured as an increase in the MOKE signal. Heat is then transferred to the Gd subsystem through spin-lattice coupling, which when combined with the AFM exchange interaction leads to a reduction of the Gd magnetic moment along the \textit{z-}direction, lowering the net magnetization. Modified exchange pathways likely also play a role in the delayed reorientation of Gd spins on these timescales. The data shows that (a) there is a delayed response of the Gd ions to the optical excitation and (b) that spin-lattice coupling and the AFM exchange interaction facilitates this.  

In both the reflectivity and MOKE signals, a clear coherent acoustic phonon is present. This strain wave launched by pump is intimately tied to the sample magnetism, with an amplitude that grows sharply at \textit{T${}_{C}$} and closely follows the magnetic order parameter. As the acoustic wave propagates it periodically alters the distance between local spins, modifying the exchange interaction. In this way, the lattice parameters are coupled to the magnetic order, which causes an attenuation of the acoustic mode near and above \textit{T${}_{C}$}, where spin fluctuations are large. This represents a laser-induced modification of the exchange interaction on ultrafast timescales through coupling to an acoustic phonon mode. While theory exists to describe magneto-elastic coupling, it is not particularly well-suited to the experiment and timescales measured here. A deeper theoretical understanding of the mechanisms at work would be instrumental in quantifying our results and motivating further studies. This work also suggests that more controlled excitation may be of interest in transiently controlling the properties of materials. An experiment of this nature has already been proposed to modify the exchange interaction in GTO, using a resonant mid-IR pulse to directly excite specific phonon modes \cite{Khalsa2018b}. The work performed here indicates the potential for GTO, and likely other titanates, as tunable magnetic materials, and highlights the need for further investigations of this nature on the road to coherent control of materials on ultrafast timescales. 

 \section{Acknowledgements}
 \label{sec:acknowledgements}
 
 We thank Leon Balents for helpful discussions and assistance with interpretation of the data. This work was supported primarily by ARO Award W911NF-16-1-0361 and additional support was provided by the W M Keck Foundation (SDW). The MRL Shared Experimental Facilities used for sample characterization are supported by the MRSEC Program of the NSF under Award No. DMR 1720256; a member of the NSF-funded Materials Research Facilities Network. The work at HYU was supported by the Basic Science Research Program through the National Research Foundation of Korea (NRF), funded by the Ministry of Science, ICT and Future Planning (2019R1A2C1084237).

\end{document}